\documentclass[a4paper,conference]{IEEEtran} 
\IEEEoverridecommandlockouts
\usepackage{cite}
\usepackage{amsmath,amssymb,amsfonts}
\usepackage{algorithmic}
\usepackage{graphicx}
\usepackage{textcomp}
\usepackage{xcolor}
\def\BibTeX{{\rm B\kern-.05em{\sc i\kern-.025em b}\kern-.08em
    T\kern-.1667em\lower.7ex\hbox{E}\kern-.125emX}}

\newcommand{\mybox}[2]{{\color{#1}\fbox{\normalcolor#2}}}
\usepackage{multirow}
\usepackage{graphicx}
\usepackage{textcomp}
\usepackage{xcolor}
\usepackage{tabularx}
\usepackage{arydshln}
\usepackage{booktabs}
\usepackage{subcaption}
\usepackage{url}
\usepackage[dvipdfmx]{hyperref}

\begin{document}

\title{Composing General Audio Representation by Fusing Multilayer Features of a Pre-trained Model}

\author{\IEEEauthorblockN{Daisuke Niizumi, Daiki Takeuchi, Yasunori Ohishi, Noboru Harada, and Kunio Kashino}
\IEEEauthorblockA{\textit{NTT Communication Science Laboratories} \\
\textit{NTT Corporation}\\
Atsugi, Japan \\
E-mail: daisuke.niizumi.dt@hco.ntt.co.jp}}

\maketitle

\begin{abstract}
Many application studies rely on audio DNN models pre-trained on a large-scale dataset as essential feature extractors, and they extract features from the last layers.
In this study, we focus on our finding that the middle layer features of existing supervised pre-trained models are more effective than the late layer features for some tasks.
We propose a simple approach to compose features effective for general-purpose applications, consisting of two steps: (1) calculating feature vectors along the time frame from middle/late layer outputs, and (2) fusing them.
This approach improves the utility of frequency and channel information in downstream processes, and combines the effectiveness of middle and late layer features for different tasks.
As a result, the feature vectors become effective for general purposes.
In the experiments using VGGish, PANNs’ CNN14, and AST on nine downstream tasks, we first show that each layer output of these models serves different tasks. Then, we demonstrate that the proposed approach significantly improves their performance and brings it to a level comparable to that of the state-of-the-art.
In particular, the performance of the non-semantic speech (NOSS) tasks greatly improves, especially on Speech commands V2 with VGGish of $+77.1$ ($14.3\%$ to $91.4\%$).
\end{abstract}


\section{Introduction}
Pre-trained models are essential building blocks as feature extractors to transfer learned representations from large-scale datasets.
In the audio domain, we can find many applications using pre-trained models: VGGish\cite{hershey2017cnn} pre-trained on YouTube-8M\cite{youtube8m} are used in conservation monitoring\cite{sethi2020soundscapes}, audio captioning\cite{koizumi2020tfmcaptioning}, and speech emotion recognition\cite{jiang2019speechemotion}; and PANNs\cite{kong2020panns} pre-trained on AudioSet\cite{gemmeke2017audioset} are used in heart sound classification\cite{koike2020heartsound} and conservation monitoring\cite{tolkova2021parsing}.

These applications use models as feature extractors without additional training, but supervised learning models are known to specialize in the pre-training dataset domain\cite{Hartmut2020WhatDo}.
On the other hand, several self-supervised learning models proposed recently show well-balanced performance on general tasks\cite{saeed2020cola}\cite{niizumi2021byol-a}\cite{wang2021universal}.
These models learn general-purpose audio representations and are considered more suitable as feature extractors.

\begin{figure}[htbp]
  \centering
  \includegraphics[width=0.9\linewidth]
  {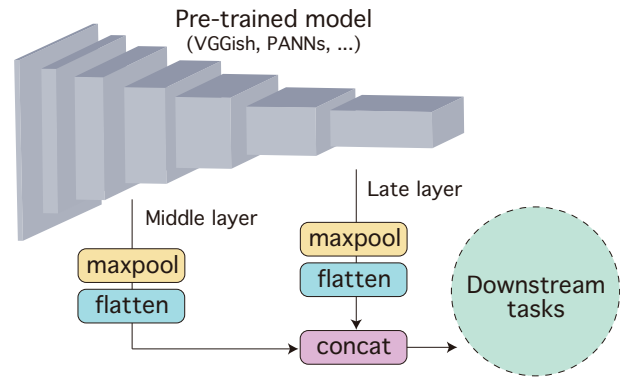}
  \caption{Proposed feature calculation flow. The number of time frames of layer outputs is adjusted by \textit{maxpool}, and the channel and frequency axes are flattened to get feature vectors along time. Then, both middle and late features are concatenated to get the final feature embeddings, making the features effective in general audio downstream tasks. The middle and late layers are chosen based on the layer-wise performance for the tasks.}
  \label{fig:concept}
  \vspace{-10pt}
\end{figure}

In this study, we focused on the potential performance of middle and late layer features of supervised pre-trained models for various tasks.
In our preliminary experiments, the late layer features of supervised pre-trained models, used in the typical applications, performed better than self-supervised pre-trained models on sound event recognition (SER) tasks (e.g., ESC-50\cite{piczak2015esc50}, UrbanSound8K\cite{salamon2014urbansound}), while they performed poorly on other tasks.
Surprisingly, however, features from the middle layer performed differently, worse on SER tasks and better on other tasks.
Our research question is: \textit{Can we put the strength of both features together for general purpose?}

We think that the reason for the imbalanced performance of the supervised learning model could be the training metric and network architecture.
Since the pre-training metric is large-dataset classification accuracy, the late layers are considered specialized to the pre-trained dataset domain\cite{Hartmut2020WhatDo}.
In addition, many of the models based on the image domain network\cite{hershey2017cnn}\cite{gong2021ast} are not designed to process time-frequency (TF) audio input effectively for audio downstream use.

To address the aforementioned problems, we propose a simple approach to calculate general-purpose features by using the outputs from the middle and late layers of a supervised pre-trained model.
In addition to the proposed approach for general-purpose feature calculation, our contributions include showing that the effective layers of the supervised pre-trained model are task-dependent and validating in experiments that the performance of these models can be significantly improved by using the proposed approach.

Our code is available at \url{https://github.com/nttcslab/composing-general-audio-repr}.

\section{Related Work}
\noindent\textbf{Feature computation of pre-trained models.}
Existing audio pre-trained models compute feature vectors similarly to how image domain models do.
VGGish\cite{hershey2017cnn} and OpenL3\cite{cramer2019openl3} flatten all axes (frequency, time, and channel) of the convolutional layer output into a single embedding vector.
COLA\cite{saeed2020cola} and TRILL\cite{shor2020trill} apply global averaging or max pooling and summarize frequency and time frame axes.

These operations can be problems for downstream tasks. For example, the pitch of the voice is considered necessary for the speaker recognition task; thus, frequency-wise information is essential. While voice inflection is vital for speech emotion recognition;  thus time-wise information is needed.

While flattening preserves all the information, it is difficult to calculate feature statistics (e.g., averaging frequency bins temporally) from flattened vectors.
On the other hand, global pooling makes information per frequency or time frame unavailable in the later processes.
Both of these issues could impair the utility of the feature vectors in downstream tasks.

\noindent\textbf{General-purpose audio representations.}
Self-supervised learning models such as COLA\cite{saeed2020cola}, BYOL-A\cite{niizumi2021byol-a}, and Slowfast NFNets\cite{wang2021universal} have been proposed for general-purpose or universal audio representations pre-trained on AudioSet without labels.
In experiments, these models generally demonstrate well-balanced performance in tasks.

\noindent\textbf{Multilevel feature fusion.}
In multimodal application research, feature fusion of multilevel (layer) outputs is utilized. For example, \cite{xu2021mffcn} fuses multilayer features from video and audio encoders.
In the image domain, \cite{chenhui2019multilayers} uses the size transformation function to match the feature size of each layer to combine multilayer features. 
In the audio domain, AudioCaps\cite{kim2019audiocaps} evaluates various audio features, including combinations of multi-layer outputs for the audio captioning task, but not for other tasks.

\section{Proposed Approach}
Our approach to improve performance of models for general tasks consists of calculating feature vectors along the time frame from layer outputs and fusing middle and late layer feature vectors.
Fig. \ref{fig:concept} illustrates the calculation flow.

To calculate feature vectors along the time frame, we first adjust the number of time frames to a $T_o$ using $\textit{maxpool}$ and then flatten the channel and frequency along time.
Adjusting the time frame of any layer feature to a $T_o$ enables subsequent \textit{fusion}, whereas flattening transforms the channel and frequency into vectors without the information loss that could be caused by averaging or max operations found in conventional methods:

\begin{equation}
\hat{z_l} = \textit{flatten}(\textit{maxpool}(z_l, T_o)), \label{eq:calc-vector}
\end{equation}
where $z_l \in R^{B\times C_l\times F_l\times T_l}$ is the $l$th layer output, and $B, C_l, F_l,$ and $T_l$ are the batch size, number of channels, number of frequency bins, and number of time frames, respectively.
The kernel and stride parameters of $\textit{maxpool}$ are set to reduce $T_l$ to $T_o$.
As a result, $\hat{z_l} \in R^{B\times C_l F_l\times T_o}$ is calculated as a feature vector with the time frame adjusted to $T_o$.

In the conventional calculation with TF features as input, flattening all axes transforms features to $R^{B\times C_l F_l T_l}$, making it difficult to use the features in later processes such as calculating the statistics of frequency along time.
Another problem is that the global averaging or max pooling transforms features to $R^{B\times C_l}$. As a result, frequency and time information is no longer available for downstream tasks.
The calculation by Eq. \eqref{eq:calc-vector} solves these problems by preserving the information for all axes and simplifying the usage of feature vectors for each time frame.

The feature vectors are fused as follows:

\begin{equation}
  z = \textit{concat}(\hat{z_M}, \hat{z_L}), \label{eq:fusion}
\end{equation}
where $\hat{z_M}$ and $\hat{z_L}$ are the features from middle layer $M$ and late layer $L$ calculated by the Eq. \eqref{eq:calc-vector}, and  $z \in R^{B\times (C_M F_M + C_L F_L)\times T_o}$ is the fused feature vector along time.
This calculation concatenates feature vectors from the middle and late layers for each time frame, which preserves all the available information from different layers.

Layer $M$ and $L$ are chosen based on the layer-wise performance for the tasks. We observed in preliminary experiments that the late layers of supervised pre-trained models excel on a set of downstream tasks $D_L$, whereas the middle layers perform better on other set of tasks $D_M$.
We choose the middle layer $M$, which shows the best average performance for $D_M$, and the late layer $L$, which shows the best average performance for $D_L$.

While $z$ provides the feature vector per time frame, the following from PANNs\cite{kong2020panns} calculates temporal statistics to make a single vector for variable-length audio.

\begin{equation}
\tilde{z} = \textit{mean}(z) + \textit{max}(z) \label{eq:mean-max-pooling}
\end{equation}

This summarizes the time axis as combined statistics of mean and max operation, and it has performed well in previous studies\cite{kong2020panns}\cite{niizumi2021byol-a}.
We used the embedding vector $\tilde{z} \in R^{B\times (C_M F_M + C_L F_L)}$ in the following experiments. 

\section{Experiments}\label{sec:experiments}
Here, we show that the performance of each layer of existing supervised learning models are task-dependent in Section \ref{sec:pre-probing}.
Next, we evaluate performance improvement of these models using our approach in Section  \ref{sec:main-eval}. Then, we compare our approach with SOTA in Section \ref{sec:sota-comparison}.

We conducted a linear evaluation using three models and nine downstream tasks.
The linear evaluation tests the effectiveness of the features of the pre-trained models by training a linear model that takes as input the features, and the test accuracy is the result.

\subsection{Experimental Details}
\noindent\textbf{Linear evaluation details.}
To train the linear model, we used the validation set for early stopping with a patience of 20 epochs and trained for up to 200 epochs with the Adam optimizer. We manually tuned the learning rate to get the best results between 0.00001 and 0.01 for every test. We ran each evaluation three times and averaged the results.

\begin{table}[htbp]
\vspace{-3pt}
\caption{VGGish layers. The ReLU layer output shape [(B)atch, (C)hannel, (T)ime, (F)requency] is calculated to [(B)atch, (D)imension, (T)ime] by Eq. \eqref{eq:calc-vector}.}
\label{tab:arch-vggish}
\centering
\begin{tabular}{cll} \toprule
Layer \# & Operation & Parameters/Output shape \\
\midrule
1 & Conv & (1, 64, kernel=(3, 3), stride=(1, 1)) \\
2 & ReLU & $[B, 64, 96, 64] \rightarrow \text{Eq. \eqref{eq:calc-vector}} \rightarrow [B, 4096, 6]$ \\
3 & MaxPool & (kernel=2, stride=2) \\
4 & Conv & (64, 128, kernel=(3, 3), stride=(1, 1)) \\
5 & ReLU & $[B, 128, 48, 32] \rightarrow \text{Eq. \eqref{eq:calc-vector}} \rightarrow [B, 4096, 6]$ \\
6 & MaxPool & (kernel=2, stride=2) \\
7 & Conv & (128, 256, kernel=(3, 3), stride=(1, 1)) \\
8 & ReLU & $[B, 256, 24, 16] \rightarrow \text{Eq. \eqref{eq:calc-vector}} \rightarrow [B, 4096, 6]$ \\
9 & Conv & (256, 256, kernel=(3, 3), stride=(1, 1)) \\
10 & ReLU & $[B, 256, 24, 16] \rightarrow \text{Eq. \eqref{eq:calc-vector}} \rightarrow [B, 4096, 6]$ \\
11 & MaxPool & (kernel=2, stride=2) \\
12 & Conv & (256, 512, kernel=(3, 3), stride=(1, 1)) \\
13 & ReLU & $[B, 512, 12, 8] \rightarrow \text{Eq. \eqref{eq:calc-vector}} \rightarrow [B, 4096, 6]$ \\
14 & Conv & (512, 512, kernel=(3, 3), stride=(1, 1)) \\
15 & ReLU & $[B, 512, 12, 8] \rightarrow \text{Eq. \eqref{eq:calc-vector}} \rightarrow [B, 4096, 6]$ \\
16 & MaxPool & (kernel=2, stride=2) \\
   & (flatten) & $[B, 12288]$\\
17 & Linear & (in=12288, out=4096) \\
18 & ReLU & $[B, 4096] \rightarrow \text{repeat} \rightarrow [B, 4096, 6]$ \\
19 & Linear & (in=4096, out=4096) \\
20 & ReLU & $[B, 4096] \rightarrow \text{repeat} \rightarrow [B, 4096, 6]$ \\
21 & Linear & (in=4096, out=128) \\
22 & ReLU & $[B, 128] \rightarrow \text{repeat} \rightarrow [B, 128, 6]$ \\
\bottomrule
\end{tabular}
\vspace{-7pt}
\end{table}

\begin{table}[htbp]
\caption{CNN14 convolutional blocks. Block output shape [B, C, T, F] is calculated to [B, D, T] by Eq. \eqref{eq:calc-vector}.}
\label{tab:arch-cnn14}
\centering
\begin{tabular}{cll} \toprule
Block \# & Output shape \\
\midrule
1 & $[B, 64, T/2, 32] \rightarrow \text{Eq. \eqref{eq:calc-vector}} \rightarrow [B, 2048, T/32]$ \\
2 & $[B, 128, T/4, 16] \rightarrow \text{Eq. \eqref{eq:calc-vector}} \rightarrow [B, 2048, T/32]$ \\
3 & $[B, 256, T/8, 8] \rightarrow \text{Eq. \eqref{eq:calc-vector}} \rightarrow [B, 2048, T/32]$ \\
4 & $[B, 512, T/16, 4] \rightarrow \text{Eq. \eqref{eq:calc-vector}} \rightarrow [B, 2048, T/32]$ \\
5 & $[B, 1024, T/32, 2] \rightarrow \text{Eq. \eqref{eq:calc-vector}} \rightarrow [B, 2048, T/32]$ \\
6 & $[B, 2048, T/32, 2] \rightarrow \text{Eq. \eqref{eq:calc-vector}} \rightarrow [B, 2048, T/32]$ \\
\bottomrule
\end{tabular}
\end{table}

\noindent\textbf{Downstream tasks.}
We employed nine tasks widely used in previous studies
\cite{kong2020panns}\cite{saeed2020cola}\cite{cramer2019openl3}\cite{shor2020trill}\cite{gong2021ast}:
ESC-50\cite{piczak2015esc50} (environmental sound classification) and UrbanSound8K (US8K, urban sound classification)\cite{salamon2014urbansound} from sound event recognition (SER) tasks;
Speech Commands V2\cite{speechcommandsv2} (SPCV2, speech command word classification), VoxCeleb1\cite{voxceleb} (VC1, speaker identification), VoxForge\cite{voxforge} (VF, language identification), and CREMA-D\cite{cao2014cremad} (CRM-D, speech emotion recognition) from non-semantic speech (NOSS) tasks; and GTZAN\cite{gt2013gtzan} (music genre recognition), NSynth\cite{nsynth2017} (music instrument family classification) and Pitch Audio Dataset (Surge synthesizer)\cite{turian2021torchsynth} (Surge, pitch audio classification) from music tasks.

\noindent\textbf{Pre-trained models.}
We tested three models: VGGish\cite{hershey2017cnn} and CNN14 from PANNs\cite{kong2020panns}, which are CNN architectures, and Audio Spectrogram Transformer (AST)\cite{gong2021ast}, which is a Transformer architecture. The followings describe their details.

\subsubsection{VGGish}
Table \ref{tab:arch-vggish} shows VGGish layers, which consists a stack of convolutional layers followed by three fully connected (FC) layers, 22 layers in total.
This model flattens all axes before the 17th layer.
Since the input time frame length is fixed to $T=96$, we converted the variable length inputs into feature vectors as follows:
encode all the divided segments of length $T$ of a input into feature vectors,
concatenate feature vectors along time,
then apply Eq. \eqref{eq:mean-max-pooling} to get a single vector for the input.
We use $T_o=6$ which is the number of time frames of 16th layer output.
The layer 18, 20, and 22 outputs don't have time axis, then we repeat them $T_o$ times to form the time axis.
We evaluate all ReLU layer outputs at layers $\in \{2,5,8,10,13,15,18,20,22\}$.

\subsubsection{PANNs' CNN14}
Table \ref{tab:arch-cnn14} shows convolutional blocks of CNN14. CNN14 accepts input with variable length $T$, and we set $T_o=T/32$. We evaluated all the block outputs.

\subsubsection{AST}
AST is a Transformer model with 12 layers, and we evaluated layer 2 to 12 outputs using 768-d {\tt [CLS]} token embeddings.
These embeddings are vectors without a time axis, unlike in CNN models.

\subsection{Evaluating Layer-wise Performance}\label{sec:pre-probing}

\begin{figure}[htbp]
  \vspace{-5pt}
  \centering
  \includegraphics[width=1.02\linewidth]
  {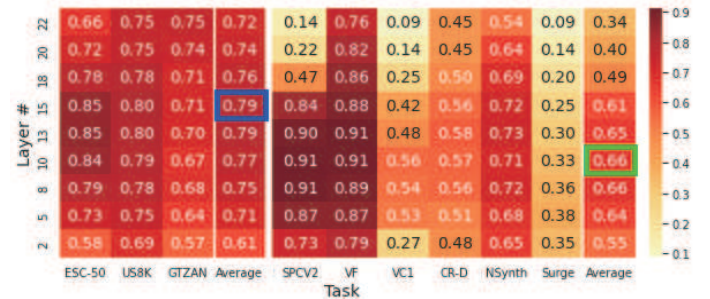}%
  \vspace{-5pt}
  \caption{VGGish layer-wise evaluation accuracies (\%).}
  \label{fig:layer-perf-vggish}
  \vspace{-10pt}
\end{figure}

\begin{figure}[htbp]
  \vspace{-5pt}
  \centering
  \includegraphics[width=1.02\linewidth]
  {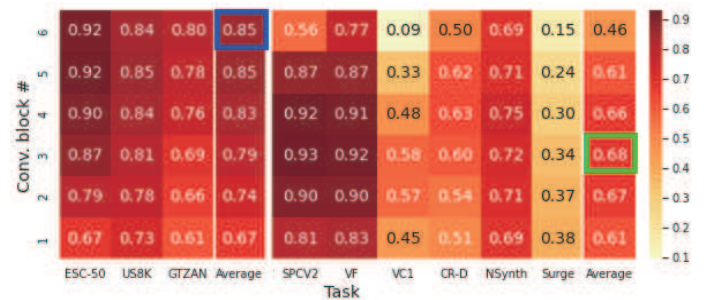}%
  \vspace{-5pt}
  \caption{PANNs' CNN14 layer-wise evaluation accuracies (\%).}
  \label{fig:layer-perf-panns}
  \vspace{-5pt}
\end{figure}

\begin{figure}[htbp]
  \vspace{-5pt}
  \centering
  \includegraphics[width=1.02\linewidth]
  {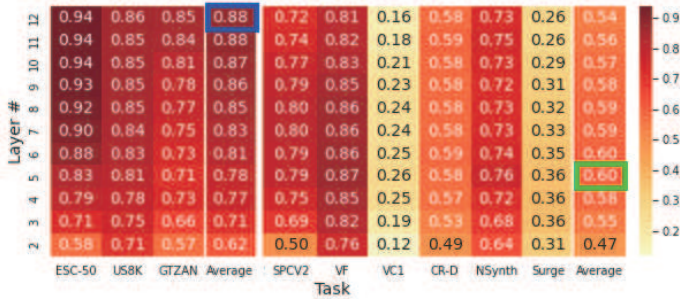}%
  \vspace{-5pt}
  \caption{AST layer-wise evaluation accuracies (\%).}
  \label{fig:layer-perf-ast}
  \vspace{-10pt}
\end{figure}

In this experiment, we evaluated the performance of each layer for all models.
The output of each layer was transformed into feature vectors per time frame using Eq. \eqref{eq:calc-vector}, and into a single vector using Eq. \eqref{eq:mean-max-pooling}.
Fig. \ref{fig:layer-perf-vggish}, \ref{fig:layer-perf-panns}, \ref{fig:layer-perf-ast} show the results for VGGish, CNN14, and AST, respectively.

\begin{table*}[htb!]
\caption{Pre-trained model accuracy improvements (\%) achieved by the proposed approach.}
\label{tab:result-main}
\centering
\begin{tabular}{lllllllllll} \toprule
&  \multicolumn{2}{c}{SER tasks} & \multicolumn{4}{c}{NOSS tasks} & \multicolumn{3}{c}{Music tasks} & \\
\cmidrule(lr){2-3} \cmidrule(lr){4-7} \cmidrule(lr){8-10}  
Representation &    ESC-50 &    US8K &    SPCV2 &    VC1 &     VF &    CRM-D &    GTZAN &     NSynth &      Surge & Avg. \\
\midrule
VGGish $^{\mathrm{1}}$ &    68.2 &    75.1 &    14.3 &     9.0 &    75.7 &    44.4 &     75.3 &    53.9 &     8.8 &    47.2 \\
VGGish-Fusion\#10\#15 &    86.5 &    80.9 &    91.4 &    54.5 &    91.6 &    59.3 &     70.8 &    73.6 &    33.3 &    71.3 \\
difference      &  +18.2 &  +5.8 &  +77.1 &  +45.5 &  +16.0 &  +15.0 &  -4.5 &  +19.7 &  +24.5 &  +24.1 \\
\midrule
CNN14 $^{\mathrm{1}}$ & 90.1 &  82.0 &    51.4 &     8.0 &    75.0 &    50.7 &     79.7 &    66.0 &    10.4 &    57.0 \\
CNN14-Fusion\#3\#6 & 93.0 &  85.8 &    91.3 &    50.6 &    90.5 &    59.0 &     77.4 &    73.8 &    32.4 &    72.6 \\
difference     & +2.9 &  +3.8 &  +39.9 &  +42.7 &  +15.5 &  +8.3 &  -2.3 &  +7.8 &  +22.0 &  +15.6 \\
\midrule
AST $^{\mathrm{1}}$ &  93.5 &  85.5 &  71.8 &  16.5 &  81.2 &  57.9 &  84.3 &  73.2 &  25.8 &  65.5 \\
AST-Fusion\#5\#12 &  94.2 &  85.5 &  80.4 &  24.9 &  87.6 &  60.7 &  82.9 &  77.6 &  34.6 &  69.8 \\
difference    &  +0.6 &  +0.0 &  +8.6 &  +8.4 &  +6.4 &  +2.8 &  -1.4 &  +4.5 &  +8.9 &  +4.3 \\
\bottomrule
\addlinespace[0.05cm]
\multicolumn{10}{l}{$^{\mathrm{1}}$
The baseline results used the last layer features from the original models.}\\
\end{tabular}
\end{table*}

The results show that the peaks of performance in a layer are different for each task. For the ESC-50/US8K/GTZAN, the peaks are in the late layers, while for the other tasks, they are in the middle layers.
This is also clearly shown by comparing the average peaks for the ESC-50/US8K/GTZAN tasks in the \mybox{blue}{blue box} with the average peak for other tasks in the \mybox{green}{green box}.

Focusing on the layer-wise results, we see that late layers perform well on ESC-50/US8K/GTZAN, while on the others, especially NOSS tasks such as VC1, they perform quite poorly, showing a substantial gap between tasks.
On the other hand, middle layers with the green box perform worse on ESC-50/US8K/GTZAN.
While no single layer satisfies all task performances, the late layers perform more imbalanced.

The VGGish results show another problem: the performance drops after layer \#15.
One possible reason is that late layers are specialized to the pre-training dataset.
Another reason could be the difference in calculation; while features up to \#15 are calculated using Eq. \eqref{eq:calc-vector}, the features after \#15 are calculated by flattening all axes.
Many VGGish application studies use the feature from late layer \#22 (FC2)\cite{sethi2020soundscapes}\cite{koizumi2020tfmcaptioning}\cite{jiang2019speechemotion}; however, other layer features calculated by using Eq. \eqref{eq:calc-vector} calculation potentially become more effective for these applications.

\begin{table*}[htb!]
\caption{Comparison with state of the art models (\%).}
\label{tab:results-sota}
\centering
\begin{tabular}{llllllllll} \toprule
 &  \multicolumn{2}{c}{SER tasks} & \multicolumn{4}{c}{NOSS tasks} & \multicolumn{3}{c}{Music tasks} \\
\cmidrule(lr){2-3} \cmidrule(lr){4-7} \cmidrule(lr){8-10} 
Representation &    ESC-50 &    US8K &    SPCV2 &    VC1 &     VF &    CRM-D &    GTZAN &     NSynth &      Surge \\
\midrule
SF-NFNet-F0\cite{wang2021universal} & 91.1 & N/A & \textbf{93.0} & \textbf{64.9} & 90.4 & N/A & N/A & \textbf{78.2} &N/A \\
COLA\cite{saeed2020cola}  & N/A & N/A & 62.4 &   29.9 &  71.3 & N/A  & N/A & 63.4 &  N/A  \\
OpenL3\cite{cramer2019openl3} $^{\mathrm{1}}$  & 79.8 & 79.3 & \underline{87.9} & \underline{40.7} & \underline{90.1} & \underline{60.4} & \underline{73.3} & \underline{75.6} & \textbf{\underline{36.4}} \\
BYOL-A\cite{niizumi2021byol-a} $^{\mathrm{1}}$ & \underline{83.7} & 79.1 & 92.2 & 40.1 & 90.2 & \textbf{\underline{62.8}} & \underline{73.6} & 74.1 & \underline{26.2} \\
\midrule
VGGish-Fusion\#10\#15 &    86.5 &    80.9 &    91.4 &    54.5 &    \textbf{91.6} &    59.3 &     70.8 &    73.6 &    33.3 \\
CNN14-Fusion\#3\#6 & 93.0 &  \textbf{85.8} &    91.3 &    50.6 &    90.5 &    59.0 &     77.4 &    73.8 &    32.4  \\
AST-Fusion\#5\#12 &  \textbf{94.2} &  85.5 &  80.4 &  24.9 &  87.6 &  60.7 &  \textbf{82.9} &  77.6 &  34.6 \\
\bottomrule
\addlinespace[0.05cm]
\multicolumn{10}{l}{$^{\mathrm{1}}$
\underline{Underlined results} were obtained in this study using publicly available pre-trained models.}\\
\end{tabular}
\end{table*}

\subsection{Evaluating Proposed Approach}\label{sec:main-eval}
In this experiment, we applied the proposed approach to VGGish, CNN14, and AST to evaluate the performance improvement.

The middle layer $M$ and late layer $L$ were determined for each model using the results of Section \ref{sec:pre-probing}.
For $L$, we choose the layer in the \mybox{blue}{blue box}, where the peaks of the average result for ESC-50/US8K/GTZAN, and for $M$, we choose the layer in the \mybox{green}{green box}, where the peaks of the average result for other tasks.
Thus, we used $M=10$ and $L=15$ for VGGish, denoting the audio representation calculated by the proposed approach as VGGish-Fusion\#10\#15.
The same goes for CNN14-Fusion\#3\#6 for PANNs' CNN14 with $M=3$ and $L=6$, and AST-Fusion\#5\#12 for AST with $M=5$ and $L=12$.

Table \ref{tab:result-main} compares the results before and after the application of the proposed approach and shows a significant improvement in performance for all models.
In particular, the performance was greatly improved in the NOSS task, especially SPCV2, which showed a significant improvement of +77.1 from 14.3\% to 91.4\% with VGGish, and improvement of +39.9 from 51.4\% to 91.3\% in CNN14.

In addition, the CNN models (VGGish, CNN14) also improved the performance of Surge (pitch classification) significantly, suggesting the contribution of local features.
The earlier the CNN layer is, the higher the frequency resolution becomes, which could make it easier to detect the pitch.

The similar performance improvements of AST in NOSS and Surge tasks show that the proposed approach is also effective for the Transformer architecture. 
As a previous study \cite{raghu2021dovision} showed that early layers attend both locally and globally,
the local feature in the fused earlier layer output possibly contributed to the improvements, similar to how the CNN layer does.

The performance of the GTZAN task slightly degraded for all models, indicating that the proposed approach can also cause degradation.
However, this degradation is small compared to the overall performance improvements; thus, we think the proposed approach is generally beneficial.

\subsection{Comparison with State of the Art}\label{sec:sota-comparison}
The results shown in Table \ref{tab:results-sota} indicate that the proposed approach brings the performance of the existing models to a level comparable with that of SOTA.
It improves the inferior NOSS task performance while maintaining the SER task performance at a higher level than that of SOTA models.

These results suggest that existing supervised pre-trained models have sufficient performance potential, which the proposed approach can exploit.
We think that the improved audio representations could generally serve various tasks.

\section{Conclusion}
In this paper, we proposed an approach to improve feature calculation of existing supervised pre-trained models without fine-tuning, and showed that the resulting features could serve as general-purpose audio representations.

The proposed approach first calculates feature vectors aligned with the time frame to improve the utility of frequency and channel information in downstream processes. Then, it fuses feature vectors from the middle and late layers, combining the effectiveness of these features for different tasks.

In the experiments using VGGish, PANNs’ CNN14, and AST on nine downstream tasks, we showed that each layer output from these models serves different tasks, and showed that the proposed approach significantly improves performance and brings it to a level comparable to that of SOTA models.
Particularly, the performance of the NOSS tasks greatly improves, especially on SPCV2 with VGGish of $+77.1$ ($14.3\%$ to $91.4\%$), while maintaining higher performance on SER tasks.

Our proposed approach provides a simple way to exploit existing supervised pre-trained models as general-purpose audio representations. 
It could make future audio application studies achieve better performance.
Our code is available online.

\bibliographystyle{IEEEtran}
\bibliography{refs}

\end{document}